\documentstyle[11pt,paspconf,epsf]{article}

\begin{document}

\title{Galaxy evolution: the effect of dark matter on the chemical
evolution of ellipticals and galaxy clusters}
\author{F. Matteucci$^{1,3}$ and B.K. Gibson$^{2}$}
\affil{ 1. Department of Astronomy, University of Trieste}
\affil{2. Mount Stromlo Observatory, Australian National University,
Weston creek Post Office, Weston, ACT 2611, Australia}
\affil{3. SISSA/ISAS, Via Beirut 2-4, I-34013 Trieste, Italy}
\begin{abstract}
In this paper we discuss the chemical evolution of elliptical galaxies
and its consequences on the evolution of the intracluster medium (ICM).
We use chemical evolution models taking into account dark matter halos
and compare the results with previous models where dark matter was not 
considered. In particular, we examine the evolution of the abundances 
of some relevant
heavy elements such as oxygen, magnesium and iron and
conclude that models including dark matter halos  and an initial
mass function (IMF) containing more massive stars than the Salpeter (1955)
IMF, better reproduce
the observed abundances of Mg and Fe both in the stellar populations 
and in the ICM (ASCA results). 
We also 
discuss the origin of gas in galaxy clusters and conclude that most of it
should have a primordial origin.
\end{abstract}


\section{Introduction}
The properties of elliptical galaxies are not easy to study since 
we can measure only their integrated properties which contain information
on several stellar populations of stars with different ages and metallicity.
Unfortunately, the effects of age and metallicity upon colors and
metallicity indices are difficult to disentangle.
Detailed models of chemical evolution and population synthesis are 
therefore necessary
to solve this degeneracy. \par
What we currently know about the properties of 
elliptical galaxies can be summarized as follows:
most of them seem to be among the oldest objects in the universe containing
old red stars,
although some recent studies (Faber et al. 1995)
seem to indicate that ellipticals are not all coeval.
This is at variance with fact that
elliptical galaxies define a fundamental plane which constrains
the mass to luminosity ratio, a quantity sensitive to the age of
stellar populations, to have a scatter less than $16\%$ for ellipticals
in rich clusters (Faber et al. 1987). 
This fact argues in favor of the ellipticals being all coeval
(Renzini 1995).\par
Ellipticals follow the well known 
color-magnitude relation in the sense that their
colors become redder with increasing luminosity and metallicity (Faber 1972).
The average metallicity of these objects appears to be solar or higher 
(Weiss et al. 1995)
thus indicating
that they have undergone a fast and intense star formation history 
perhaps interrupted by the occurrence of galactic winds.
\par
Finally, another constraint to models of chemical evolution of ellipticals,
arguing for them to be old systems, is the measured SN rate. These galaxies
have only SNe of Type Ia which are believed to originate from 
old stars, namely white dwarfs in binary systems.
\par
Metallicity indeces of Fe and Mg seem to indicate 
an overabundance of $\alpha$-elements relative to iron in the nuclei of
elliptical galaxies and that this overabundance increases with
galactic luminosity. On the other hand, the Mg/Fe ratio seems to be
constant inside the same galaxy (Worthey et al. 1992; Carollo and Danziger
1994), thus indicating that the gradients of Mg and Fe inside the galaxies
are roughly the same. The existence of such gradients seems to be now
well established (Carollo et al. 1993) and clearly indicates that some
dissipative processes have occurred during the formation of ellipticals.
\par
Elliptical galaxies are also known to have large halos of gas 
emitting in the X-ray band.  These halos represent an indirect indication
that these galaxies should contain extended dark matter halos.
Other indications of the existence of such halos comes from HI kinematics, 
planetary nebulae, gravitational lensing and dynamical studies 
(see Carollo et al.
1995 and references therein).  
The abundances measured  by ASCA
in the X-ray emmitting gas indicate a very low iron
abundance in disagreement with the results from stellar populations.
However, the derivation of abundances from the X-ray spectra 
is still affected by many uncertainties, as recently discussed by 
Arimoto et al. (1996).
Another result from ASCA, influencing the evolution of ellipticals, concerns
the abundances derived for the ICM in galaxy clusters. They
indicate a clear overabundance of $\alpha$-elements relative to iron
in the ICM (Mushotzky 1994).
This result is relevant to 
the chemical evolution
of elliptical galaxies, since they are the major contributors to the enrichment
of the ICM, as we will see in the next sections.\par

\section{The standard model of chemical evolution}

The model we are going to discuss has been developed 
first in Matteucci and Tornamb\`e (1987) and
subsequently refined in Matteucci (1992), Matteucci and Gibson
(1995) 
and Gibson and Matteucci (1997)
where a detailed description can be found. We will only remind here
that it belongs to the category of the supernovae (SNe)-driven 
galactic wind models.
In this model, efficient star formation leads to the
development of a galactic wind, once the thermal energy of the gas, 
due to the energy injection from SNe and stellar winds, is equal or larger 
than the binding energy of the gas.
The star formation is assumed to stop after the occurrence of the galactic wind
and the galaxy evolves passively thereafter. During this stage the only
active process is the restoration of gas from the stars into 
the interstellar medium (ISM).
The only active SNe in this phase are the Type Ia which continue 
exploding until the present time in agreement with the observational evidence.
The model includes the most recent ideas on SN progenitors and nucleosynthesis,
indicating that SNe Ia originate from long living stars whereas SNe of Type II
originate from short living stars. Type Ia SNe are believed to produce
roughly $\simeq 0.6 M_{\odot}$ of Fe, and the SNe of Type II roughly
$\simeq 0.1 M_{\odot}$ of Fe. The so-called $\alpha$-elements 
(O, Mg, Ne, Si, etc..) are thought to be preferentially formed in Type II SNe.
\par
\subsection{Basic equations}
We follow the evolution of the abundances of 13 chemical elements as 
described in Matteucci and Gibson (1995).
\par
The initial mass function $\varphi(m)$ is assumed to be
constant in time and expressed as a power law. We explore 
several prescriptions for the IMF:
a) Salpeter (1955), b) Arimoto and Yoshii (1987) and c) Kroupa et al. (1993).
\par
The star formation rate is assumed to be simply proportional to the 
volume gas density through a constant, the star formation efficiency. 
This efficiency
is constant in time but is assumed to vary with the galactic mass in the 
sense that
the efficiency decreases when the total mass increases. This is based 
on a suggestion from Arimoto and Yoshii (1987) that the efficiency of star 
formation should be inversely proportional to the dynamical timescales.
This assumption is not necessarily true since we could think that the 
star formation efficiency may increase with galactic mass as suggested
by Tinsley and Larson (1979) and Matteucci (1994) in the hypothesis of
the formation of ellipticals by merging of gaseous fragments.
According to these different assumptions on the star formation efficiency 
we can predict very different behaviours of the abundances and 
abundance ratios as
functions of galactic mass, as we will see in the following. 
\par
The binding energy of the gas is computed by assuming that elliptical
galaxies possess heavy but diffuse halos of dark matter and we follow the
formulation of Bertin et al. (1992). According to these authors
and Matteucci (1992) we can write the binding energy of the gas as the 
sum of two terms, one being the binding energy of gas due to the luminous 
matter and the other the binding energy of the gas due to the interaction between dark and luminous matter:
\begin{eqnarray}
W=W_L+W_{LD}
\end{eqnarray}

\begin{eqnarray}
W_L=-0.5G{M_{gas}(t)M_{lum}(t) \over r_L}
\end{eqnarray}

\begin{eqnarray}
W_{LD}=-G{M_{gas}(t)M_{dark}(t) \over r_L}\tilde W_{LD}
\end{eqnarray}
\begin{eqnarray}
\tilde W_{LD}\simeq {1 \over 2\pi}{r_L \over r_D}[1+1.37({r_L \over r_D})]
\end{eqnarray}
where $M_{gas}$, $M_{lum}$ and $M_{dark}$ are the mass of gas, the mass
of luminous matter and the mass of the dark matter, respectively.
The quantities $r_L$ and $r_{D}$ are the half-light radius and the the
radius of the dark matter core, respectively. In the Bertin et al. (1992) formulation the ratio ${r_L \over r_D}$ can vary in the interval 0.1 $\rightarrow$
0.45. The case with dark matter distributed like luminous matter, 
although unrealistic, is represented by $\tilde W_{LD}=1$.

The total mass of the galaxy is defined as:
\begin{eqnarray}
M_{tot}=M_{lum}+M_{dark}
\end{eqnarray}
which can be written as:
\begin{eqnarray}
M_{tot}=(1+R)M_{lum}
\end{eqnarray}
where $R={M_{dark} \over M_{lum}}$ is the ratio between the mass of the
dark and luminous component.
\par
For the sake of simplicity
we assume that every galaxy, irrespective of its luminous mass, has $R=10$ and
${r_L \over r_D}=0.1$. This means heavy but diffuse halos of dark matter.
The choice of relatively diffused dark
matter halos seems to be appropriate for elliptical galaxies as shown by 
the models Matteucci (1992) and Matteucci and Gibson (1995) as well as by the
lack of dynamical evidence for dark matter inside one or two optical radii
(Carollo et al. 1995).
However, we do not exclude a possible variation of the amount 
and/or concentration of dark matter with galactic luminosity, 
as it seems to be the
case in spiral galaxies (Persic and Salucci 1988; Persic et al. 1996).
The possibility of such variations has been already discussed by 
Renzini and Ciotti (1993) in order to explain the properties of the fundamental
plane for ellipticals.
\par
The thermal energy of the gas is calculated by considering both the SN
and stellar wind energy injection.
The contribution to the thermal energy of gas from SNe is written as:
\begin{eqnarray}
E_{th_{SN}}=\int_0^{t}{\epsilon(t-t^{'})R_{SN}(t^{\prime\prime})dt^{'}}
\end{eqnarray}
where $R_{SN}(t)$ is the SN rate (either Type I or II) and 
$\epsilon_{SN}(t_)$ is the fraction of the initial blast wave energy
which is transferred into the ISM as thermal energy and $t^{'}$ 
is the explosion time.
For the particular form of $\epsilon_{SN}(t)$ see Matteucci and Gibson (1995)
and Gibson and Matteucci (1997).
\par
The contribution to the thermal energy of the gas from stellar 
wind is calculated
according to Gibson (1994) and is written as:
\begin{eqnarray}
E_{th_{SW}}=\eta\int_0^{t}\int_{12.0}^{M_U}{{\varphi(m)
\over m} \epsilon_W(m,t-t^{'},Z(t^{'}))dm dt^{'}}
\end{eqnarray}
\par
When the total thermal energy of gas is equal or larger than its binding
energy, a galactic wind develops and lasts until this condition is 
reversed.
Whether there is only an early wind episode or whether more episodes 
occur is a delicate point and depends
crucially on the balance between the effects of the potential well and 
the injection of energy from stars.
Clearly the presence and distribution of dark matter plays a very 
important role in determining the onset and the entity of galactic winds.

\section{Results for elliptical galaxies}

For standard model described herein the star formation efficiency decreases
with galactic mass.  This model
predicts that more massive galaxies develop a galactic wind later than 
the less massive ones, the reason being that the potential well depth
increases
with the total mass of the galaxy, while the efficiency of star formation 
decreases. This behaviour can change if one assumes that the efficiency of 
star formation is increasing with the total mass and one can obtain 
the situation where the more massive galaxies develop a galactic wind before 
the less massive ones. Matteucci (1994) analysed this case and referred to it
as the inverse wind scenario.
The reason for requiring 
an inverse wind situation resides in the fact that one cannot explain the 
increase of the [Mg/Fe] ratio 
in the nuclei of ellipticals as a function of galactic luminosity
(Worthey et al. 1992). In fact, the standard model predicts exactly 
the opposite behaviour, due to the winds developing later in more massive 
ellipticals, which drives the average
[Mg/Fe] downward. This ``downward'' trend results from the increased
contribution of Fe-donating Type Ia SNe in the more massive galaxies.
This behaviour is well illustrated in Figure 1 where we show the predictions
of the standard model concerning the [O/Fe] ratio (oxygen and magnesium
should vary in lockstep) in the gas for model computed with different IMFs
and different initial luminous masses. The times for the occurrence of the
galactic wind is marked on each curve.
From this figure it is easy to see that if the times for the 
occurrence of the galactic wind become shorter and shorter with 
increasing galactic 
mass, then the average
[O/Fe] would increase  with galactic mass instead than decrease.
Another possibility, in order to obtain an inverse wind situation, 
might be to vary the amount and/or the concentration of dark 
matter in a way such that the
more massive galaxies would have less and/or less concentrated dark matter.
This sounds like an interesting possibility, although it has
not yet been calculated in detail, since it seems to happen 
in spiral galaxies (Persic and Salucci 1988; Persic et al. 1996)
and in dwarf spheroidal galaxies (Kormendy, 1990). 
From this discussion it becomes evident how abundance ratios in 
stellar populations and gas in
ellipticals can be used to constrain the amount and concentration 
of dark matter in these objects.
Another possibility, however, could be a variable IMF from galaxy to galaxy.
In this case, in fact, one does not need to have an inverse wind situation, 
as shown in Matteucci (1994).
This possibility has also the advantage of reproducing the slight 
increase of the $M/L_B$ ratio with galactic mass (Bender et al. 1992).
The variation of the IMF should be such that more massive galaxies should have
more massive stars relatively to less massive galaxies.

\begin{figure}
\plotone{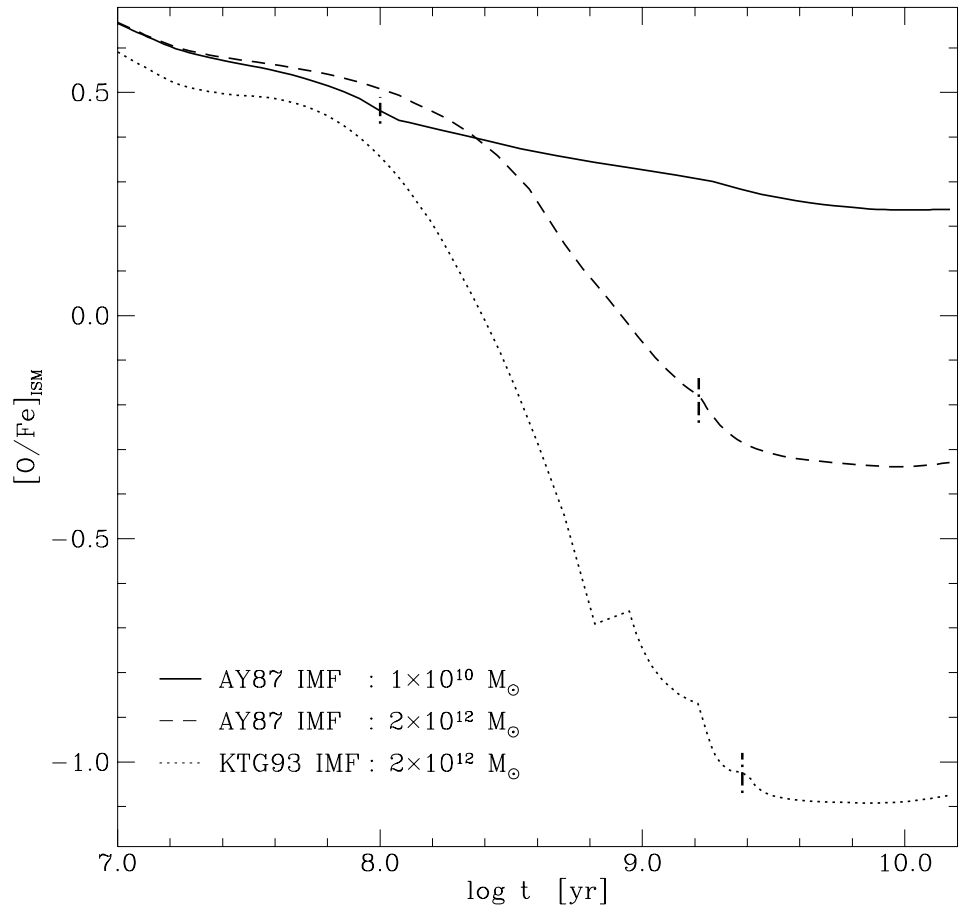}
\caption{The [O/Fe] ratio in the gas as a function of time for
galaxies of different initial luminous mass and different IMF's prescriptions:
AY87=Arimoto and Yoshii (1987), KTG93=Kroupa et al. (1993)} \label{fig-1}
\end{figure}

Finally,
a very important feature of Figure 1 is that, independently of the assumed 
IMF and time for the occurrence of a galactic wind, a high
[O/Fe] ratio in the dominant stellar population is achieved only if the 
period of major star formation has been short, namely no longer than $
\sim3 \cdot 10^{8}$
years. This is a very strong conclusion since it implies a very fast process
for the formation of big ellipticals, at variance with   
the hierarchical clustering scenario for galaxy formation.

\par
Matteucci and Gibson (1995) calculated several models for ellipticals 
of initial luminous mass in the range 
$1.0 \cdot 10^{9} \rightarrow 2.0 \cdot 10^{12} M_{\odot}$ for the 
three IMF cases a), b) and c). One set of models was calculated 
with an increasing 
star formation efficiency. Their results are shown in Table 1.

\small
\begin{table}
\caption {Predicted masses of gas and individual elements ejected
by a global wind at time $t_{\rm GW}$ from a galaxy of 
initial mass $M_{\rm g}(t\equiv 0)$.
Columns 8 and 9 contain the predicted present-day
SNe Type Ia rates (SNe century$^{-1}$ $10^{-10}$L$_{B_\odot}$)
and luminous mass to blue luminosity ratios, respectively.  
The mass-weighted mean stellar metallicity is given in column 10.} 
\label{tbl-1}
\begin{center}\scriptsize
\begin{tabular}{cccccccccc}\hline
\small
$M_{\rm g}(0)$ & $t_{\rm GW}$ & 
$M_{\rm g}^{\rm ej}$ &
$M_{\rm Fe}^{\rm ej}$ &
$M_{\rm O}^{\rm ej}$ &
$M_{\rm Mg}^{\rm ej}$ &
$M_{\rm Si}^{\rm ej}$ &
$R_{\rm SNIa}$ & $M_{\rm lum}/L_{\rm B}$ & [Fe/H]$_m$ \\ \hline
\multicolumn{10}{c}{Salpeter IMF: Inverse Wind Model ($A=0.18$)} \\
1.0(9)  & 2.21 & 4.50(7)  &  6.74(5) &  1.46(6) &  1.00(5) &  2.38(5) & 0.32 & 13.8 & +0.11 \\
1.0(10) & 1.16 & 3.71(8)  &  4.40(6) &  1.12(7) &  7.11(5) &  1.60(6) & 0.34 & 14.9 & -0.03 \\
1.0(11) & 0.56 & 1.80(9)  &  2.37(7) &  5.95(7) &  3.84(6) &  8.68(6) & 0.23 & 15.6 & -0.09 \\
1.0(12) & 0.24 & 1.59(10) &  1.33(8) &  2.38(8) &  1.70(7) &  4.73(7) & 0.08 & 15.9 & -0.28 \\
2.0(12) & 0.19 & 2.94(10) &  2.36(8) &  5.57(8) &  3.65(7) &  8.96(7) & 0.02 & 16.0 & -0.28 \\ \hline
\multicolumn{10}{c}{Salpeter IMF: Classic Wind Model ($A=0.18$)} \\
1.0(9)  & 0.06 & 3.48(8)  &  2.89(5) &  5.50(6) &  2.24(5) &  2.45(5) & 0.00 & 16.1 & -0.78 \\
1.0(10) & 0.20 & 9.43(8)  &  3.29(6) &  2.97(7) &  1.46(6) &  1.95(6) & 0.09 & 16.0 & -0.33 \\
1.0(11) & 0.56 & 1.85(9)  &  2.30(7) &  6.86(7) &  4.24(6) &  8.75(6) & 0.26 & 15.6 & -0.03 \\
1.0(12) & 1.34 & 6.46(9)  &  1.40(8) &  2.20(8) &  1.73(7) &  4.66(7) & 0.30 & 14.8 & +0.17 \\
2.0(12) & 1.80 & 9.79(9)  &  2.90(8) &  3.40(8) &  3.06(7) &  9.24(7) & 0.29 & 14.2 & +0.26 \\ \hline
\multicolumn{10}{c}{Arimoto \& Yoshii IMF: Classic Wind Model ($A=0.05$)} \\
1.0(9)  & 0.02 & 7.17(8)  &  5.15(5) &  1.23(7) &  5.04(5) &  4.79(5) & 0.00 & 23.6 & -0.60 \\
1.0(10) & 0.10 & 4.22(9)  &  1.15(7) &  2.13(8) &  1.04(7) &  1.04(7) & 0.00 & 23.5 & -0.20 \\
1.0(11) & 0.55 & 6.57(9)  &  7.03(7) &  5.09(8) &  4.78(7) &  5.26(7) & 0.27 & 22.7 & +0.29 \\
1.0(12) & 1.28 & 2.16(10) &  3.37(8) &  1.56(9) &  1.78(8) &  2.14(8) & 0.31 & 21.4 & +0.41 \\
2.0(12) & 1.64 & 3.14(10) &  5.24(8) &  2.22(9) &  2.59(8) &  3.18(8) & 0.29 & 20.8 & +0.43 \\ \hline
\multicolumn{10}{c}{Arimoto \& Yoshii IMF: Classic Wind Model ($A=0.02$)} \\
1.0(9)  & 0.02 & 7.17(8)  &  5.15(5) &  1.23(7) &  5.04(5) &  4.79(5) & 0.00 & 23.6 & -0.60 \\
1.0(10) & 0.10 & 4.22(9)  &  1.15(7) &  2.13(8) &  1.04(7) &  1.04(7) & 0.00 & 23.5 & -0.20 \\
1.0(11) & 0.55 & 6.57(9)  &  7.03(7) &  5.09(8) &  4.78(7) &  5.10(7) & 0.09 & 22.7 & +0.25 \\
1.0(12) & 1.28 & 2.16(10) &  3.37(8) &  1.56(9) &  1.78(8) &  1.73(8) & 0.11 & 21.3 & +0.34 \\
2.0(12) & 1.64 & 3.14(10) &  5.24(8) &  2.22(9) &  2.59(8) &  2.66(8) & 0.10 & 20.7 & +0.36 \\ \hline
\multicolumn{10}{c}{Kroupa et al. IMF: Classic Wind Model ($A=0.30$)} \\
1.0(9)  & 0.10 & 1.83(8)  &  2.02(5) &  2.39(6) &  9.79(4) &  1.31(5) & 0.00 & 8.3 & -0.79 \\
1.0(10) & 0.24 & 5.83(8)  &  2.91(6) &  1.18(7) &  6.02(5) &  1.19(6) & 0.14 & 8.3 & -0.39 \\
1.0(11) & 0.70 & 1.32(9)  &  2.41(7) &  2.89(7) &  2.29(6) &  7.49(6) & 0.20 & 8.0 & +0.04 \\
1.0(12) & 1.87 & 6.15(9)  &  2.43(8) &  1.60(8) &  1.78(7) &  7.14(7) & 0.25 & 7.2 & +0.31 \\
2.0(12) & 2.40 & 1.07(10) &  4.77(8) &  2.89(8) &  3.40(7) &  1.39(8) & 0.28 & 6.9 & +0.39 \\ \hline

\end{tabular}
\end{center}
\noindent
\end{table}
\normalsize

All of these models contain the same prescriptions about dark matter. 
The interesting fact is that in all the models they found only early 
galactic winds and
this was mainly due to the presence of dark matter. The epoch for 
the onset of the galactic winds is shown in column 2, the duration 
of the wind phases 
varies from several $10^{7}$ years to several $10^{8}$ years. After the wind
the galaxies evolve passively just accumulating the gas restored by all the
dying stellar populations. In column 3 is shown the total mass 
of gas which is ejected during the wind phase, while in column 
4,5,6 and 7 are shown the
masses ejected in the form of Fe,O,Mg and Si. In column 8 we show 
the present time Type Ia SN rate in units of SNu which should be compared 
with the observed rate: 
$(0.25 \rightarrow 0.44)h^{2}$ SNu (Turatto et al. 1994), where $h=H_o/100$.
Finally in column 9 we show the predicted $M/L_B$ ratios and in column 10
the predicted mass-weighted 
average $<[Fe/H]>$ in the dominant stellar population

\section{Contribution of the ellipticals to Fe and $\alpha$-elements 
of the ICM}
Matteucci and Gibson (1995) derived power relationships 
between the ejected masses $M_i^{ej}$ of gas/chemical species 
and the final galactic mass from the results shown in Table 1.
Then they integrated these ejected masses on the cluster mass spectrum
(Schechter 1976)
under the assumption that mainly ellipticals and S0 galaxies contribute
ejected gas to the ICM. 
This assumption is generally supported by the observational evidence that 
there is a clear correlation between the total visual luminosity
of the ellipticals in clusters and the total masses of iron and gas measured
in the X-ray band (Arnaud 1994).
However, as we will see in the next section, there could be the 
possibility that dwarf galaxies, previously underestimated in clusters, 
contribute non-negligibly to the total gas mass in clusters. 
The derived integrated masses together with the predicted 
[O/Fe], [Si/Fe] and iron mass-to-light ratio for the clusters (IMLR) are shown
in Table 2. Table 2 shows the results obtained for a typical rich and
poor cluster. All the models are computed by assuming 
$H_{o}=85$ km sec$^{-1}$ Mpc$^{-1}$ with the exception 
of the classic wind model with the Arimoto and Yoshii (1987) IMF where
the results for $H_{o}=50$ km sec$^{-1}$ Mpc$^{-1}$ (second row of each case)
are also shown.
\small
\begin{table}
\caption{
Predicted total mass of gas and elements ejected into the
intracluster medium (ICM) from all cluster ellipticals and lenticulars, and 
ICM Iron Mass-to-Light Ratios (IMLRs), [O/Fe] and [Si/Fe] ratios.} 
\label{tbl-2}  
\begin{center}\scriptsize
\begin{tabular}{cccccccc}\hline
\small
$M_{\rm g,tot}$ & $M_{\rm Fe,tot}$ & 
$M_{\rm O,tot}$ & $M_{\rm Mg,tot}$ & $M_{\rm Si,tot}$ &
IMLR & [O/Fe] & [Si/Fe] \\ \hline
\multicolumn{7}{c}{Salpeter IMF: Inverse Wind Model} \\
\multicolumn{7}{c}{\it Rich Cluster} \\
3.83(12)& 3.81(10)& 8.60(10)& 5.34(9) & 2.16(10) & 0.002 & -0.46 & -0.17 \\
\multicolumn{7}{c}{\it Poor Cluster} \\
1.13(11)& 1.17(9) & 2.64(9) & 1.64(8) & 6.57(8)  & 0.002 & -0.45 & -0.18 \\
\hline
\multicolumn{7}{c}{Salpeter IMF: Classic Wind Model} \\
\multicolumn{7}{c}{\it Rich Cluster} \\
4.76(12)& 4.04(10)& 1.10(11)& 6.90(9) & 1.94(10) & 0.002 & -0.37 & -0.25 \\
\multicolumn{7}{c}{\it Poor Cluster} \\
1.68(11)& 1.19(9) & 3.77(9) & 2.28(8) & 6.05(8)  & 0.003 & -0.30 & -0.22 \\
\hline
\multicolumn{7}{c}{Arimoto \& Yoshii IMF: Classic Wind Model} \\
\multicolumn{7}{c}{\it Rich Cluster} \\
1.85(13)& 1.10(11)& 8.59(11)& 7.07(10)& 1.15(11) & 0.006 & +0.08 & +0.09 \\
1.44(13)& 4.18(10)& 5.52(11)& 3.26(10)& 3.64(10) & 0.005 & +0.31 & +0.19 \\
\multicolumn{7}{c}{\it Poor Cluster} \\
6.57(11)& 3.38(9) & 2.88(10)& 2.24(9) & 3.60(9)  & 0.006 & +0.12 & +0.10 \\
1.03(12)& 2.62(9) & 3.72(10)& 2.08(9) & 2.30(9)  & 0.006 & +0.34 & +0.19 \\
\hline
\multicolumn{7}{c}{Kroupa et al. IMF: Classic Wind Model} \\
\multicolumn{7}{c}{\it Rich Cluster} \\
2.37(12)& 4.12(10)& 5.17(10)& 3.66(9) & 1.48(10) & 0.003 & -0.71 & -0.38 \\
\multicolumn{7}{c}{\it Poor Cluster} \\
8.07(10)& 1.13(9) & 1.70(9) & 1.13(8) & 4.26(8)  & 0.003 & -0.63 & -0.36 \\
\hline
\end{tabular}
\end{center}
\end{table}
\normalsize

One of the
most relevant results shown in Table 2 is the total mass of 
gas ejected from the ellipticals which is,
in all cases, far less than observed indicating that the majority 
of the ICM should have a primordial origin. This conclusion had 
already been reached by Matteucci and Vettolani (1988) but in 
this case the discrepancy with the observed total gas masses is even 
stronger since, due to the
presence of dark matter halos, 
the galaxies do not eject all of their available gas as it was the case in the
Matteucci and Vettolani (1988) models. 
Therefore, 
the presence of dark matter in galaxies has important consequences also for
the interpretation of the origin of the ICM as we will see even 
better in the next section.
Another important result is the predicted [O/Fe] ratio in the ICM. 
Recent ASCA data suggest that [O/Fe] is high and positive 
($0.1 \rightarrow 0.7$,
Mushotzky 1994). From Table 2 we can see that only models with
a flat IMF (case b) can reproduce the observed values. It is worth 
noting that the existence of only early winds is also a necessary, although not 
sufficient, condition to achieve this situation.
In fact, if all the gas restored by stars is allowed to eventually be lost,
the resulting [O/Fe] ratio in the ICM is low and negative as a result 
of the continuous injection of Fe from SNe of Type Ia, as it was 
the case in Matteucci and Vettolani(1988)'s model.
Abundance ratios in the ICM
are therefore an extremely useful and tight constraint to
understand the evolution of ellipticals. However, before drawing 
firm conclusions we should be confident of the abundances derived 
from the X-ray 
spectra, whereas the situation is not yet clear as already 
mentioned before. 

\section{More on the source of the ICM gas}

While the bright and intermediate part of the luminosity function of a
cluster is consistent with a slope $\alpha \simeq -1.25 \rightarrow
-1.45$ (Ferguson and Sandage, 1995), there are recent indications
that the faint-end slope ($M_{B} \ge -15.0$) of the luminosity 
function may be significantly steeper ($\alpha \sim -1.8 \rightarrow -2.2$,
De Propris et al. 1995). Trentham (1994) suggested that, if the faint-end 
slope of the luminosity function
is taken into account, 
we can explain all the gas in clusters as due to galaxies 
and in particular to dwarfs. 

\begin{figure}
\plottwo{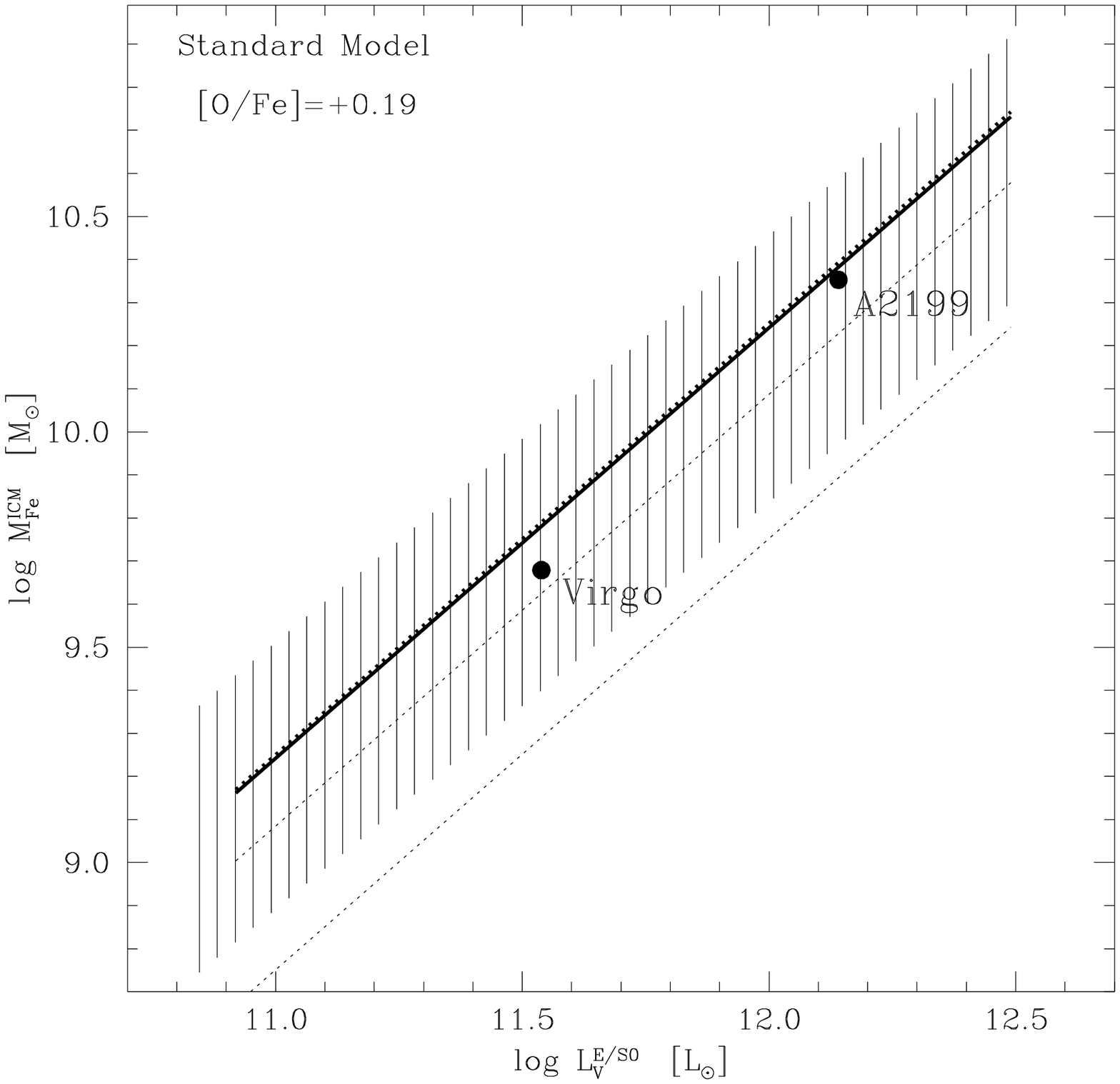}{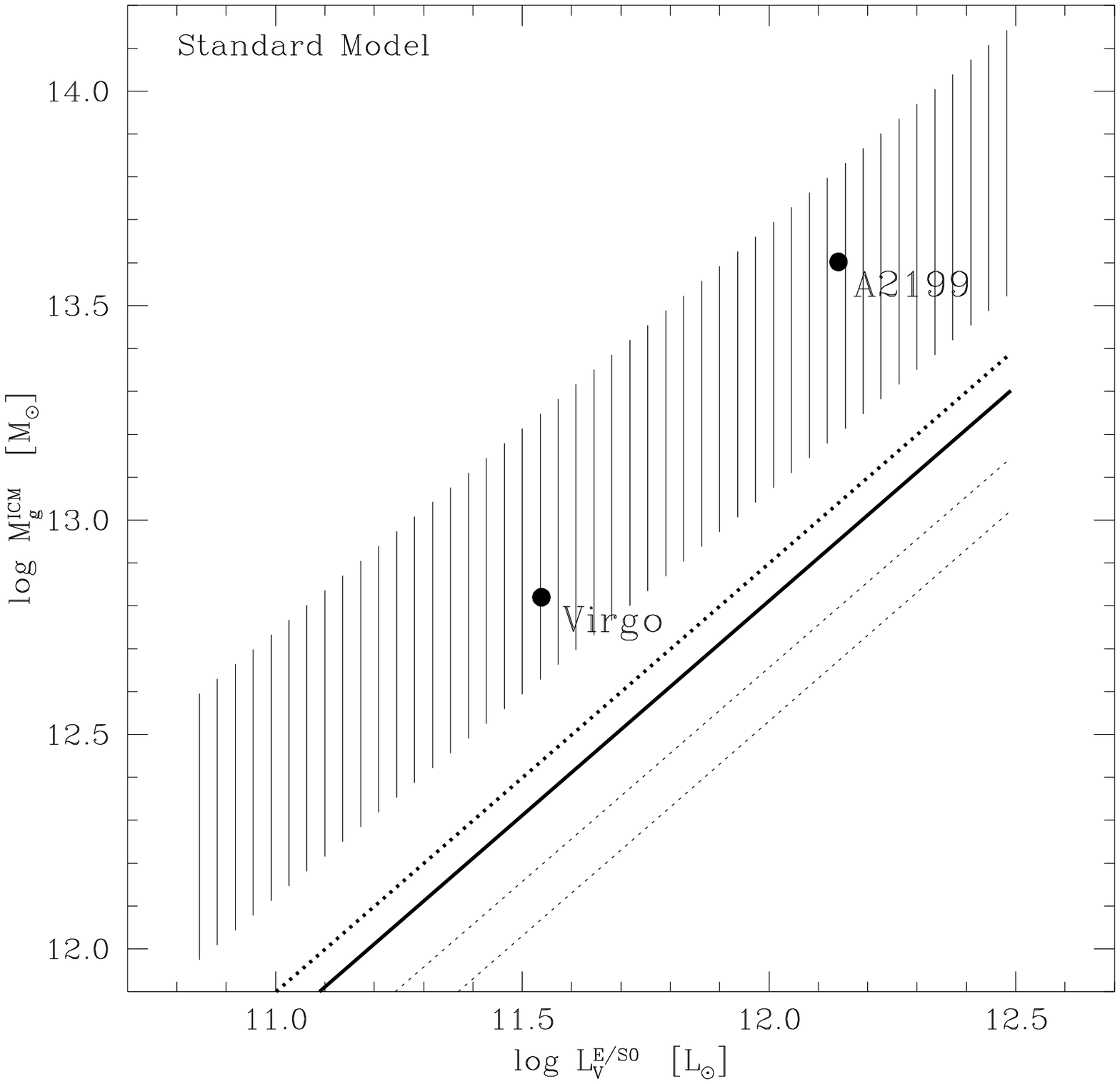}
\caption{Shaded region shows the observed correlation between the non-spiral-
originating V-band cluster luminosity, and the observed ICM iron mass
(left figure) and 
gas mass (right figure)
after Arnaud (1994). Solid curve is a single luminosity function
model of slope $\alpha=-1.45$. Dotted lines are the components of 
a two slope luminosity function model-the 
lower curve is the low luminosity dwarf
spheroidal component with $\alpha=-1.9$. The middle one is the normal giant
spheroidal population  with $\alpha=-1.45$. 
The heavy dotted curve is their sum.}
\label{fig-2}
\end{figure}

In particular, in order to achieve this,
one has to assume that each galaxy, irrespective of its mass, looses
a fraction of its total mass (luminous+dark) of the order of $\gamma=0.33$.
The range of galactic masses contributing to the gas being $10^{4}-10^{11}
M_{\odot}$ and the slope of the faint- end of the luminosity function being
in the range $\alpha=-1.4 \rightarrow -1.7$. 
Under these prescriptions he was able to 
account for all the gas in clusters as due to galaxies, but he predicted
for dwarf galaxies
$M/L \propto L^{-0.42 \rightarrow -0.27}$ not in very good agreement with
the observational estimate. Kormendy (1990), in fact, suggests that
the exponent of this law for dwarf galaxies is $\beta \sim -0.13 \rightarrow
-0.31$. This exponent $\beta$ measures how rapidly 
dwarf galaxies become dark matter dominated as their luminosity decreases. 
Later, Nath and Chiba (1995) calculated more realistic $\gamma$ values
varying with galactic mass and found that in order to explain 
all the gas in clusters as due to dwarfs one needs to assume 
$\alpha \sim -1.7 \rightarrow -1.9$,
thus obtaining a $\beta \sim -0.55 \rightarrow -0.37$, totally outside the
observed range. Therefore, they showed that more realistic assumptions 
weaken the argument of Trentham (1994).
Recently, Gibson and Matteucci (1997) extended the model 
for elliptical galaxies described before to dwarf spheroidal 
galaxies, included updated physical inputs and calculated the total
contribution to the gas and metal of the ICM from dwarf and giant elliptical
galaxies. These authors calculated also the photometric evolution 
(see Gibson 1996 for details) of the
considered galaxies in order to have a larger number of constraints 
to compare with
the results and took into account the extreme case where all the gas 
produced by stars, after the early wind phase, is eventually lost.
They found that the simple suggestion of Trentham (1994) about the 
amount of mass which should be ejected by each galaxy is extremely unrealistic.
In order to reproduce realistic galaxies, namely with the right 
colors and luminosities, they found $\gamma \sim 0.04 \rightarrow 0.08$. 
They used a slope
for the faint end of the luminosity function  $\alpha$=-1.9. 
Their results indicate that also under these extreme conditions and 
for any reasonable choice of
the main parameters is not possible to explain all of the gas in 
clusters as due to galaxies. In particular, they found that the contribution 
of dwarfs to the total gas is not negligible and it raises 
the galactic contribution up to
$35\%$ to compare with the negligible amount
of galactic gas obtained by Matteucci and Gibson (1995)
by considering only normal ellipticals. 
In Figures 2 and 3 we show the predicted total masses of iron and gas as
functions of the total visual luminosity of the cluster galaxies compared 
with the data from Arnaud (1994).
From these figures it is clear that the contribution of dwarfs 
to the gas in not negligible whereas their contribution to 
metals is negligible.
This conclusion is in agreement with Nath and Chiba (1995) who also 
showed that dwarfs contribute negligibly to the iron content in clusters.
\par
Finally, we would like to stress the fact that this result about 
the origin of gas in clusters is quite robust.
In fact, Gibson and Matteucci (1997) have chosen the most favorable 
conditions for the galaxies to loose mass. In particular, they assumed 
that the amount and concentration of dark matter in dwarfs is constant 
thus underestimating their potential well. The $\beta$ value that they 
predict is, in fact, $\beta=
-0.07$, lower than the observed range.
\par

\end{document}